# Joint SLA Admission and Risk Allocation for Cloud Resource Provisioning

Yihua Xu*†

Independent Researcher, yxu442@gmail.com

Jize Li†

Independent Researcher, jizel@bu.edu

Yan Lyu

Independent Researcher, lyu.yan@northeastern.edu

Jiani He

Independent Researcher, jianihe@alum.mit.edu

You Lin

Independent Researcher, yl2356@cornell.edu

Shangjing Tang

Independent Researcher, st787@cornell.edu

Dishu Yang

Independent Researcher, yang.dis@northeastern.edu

Cloud controllers must reserve enough CPU and memory to meet demand without leaving too much capacity idle. Under a shared service-level agreement (SLA), a shortage of CPU or memory (considered as resource) counts toward the same limit. Dividing the allowed shortage risk between them changes how much extra capacity each need. We study this choice when forecast-error histories are short or correlated over time. Our controller compares different approaches to split the allowed shortage risk between resources. For each split, the controller uses historical forecast errors to determine how much extra capacity to reserve. It checks whether the history is enough to support that split and its consistency across different parts of the history. Among the splits that pass, it chooses the one with the lowest combined cost of shortages and idle capacity. Otherwise, the controller continues serving the workload with a calibrated fallback. In our simulation, we set a joint SLA target of sufficient CPU and memory in at least 95% of time slots and used 192 hours of historical forecast errors to structure the controller. Compared with always using fallback, the controller reduced the time slots allocation with either of the resource shortage by 1.04 percentage points on average. This reduction came with a 0.0091 increase in normalized cost from shortages and idle capacity. In the Bitbrains trace, five nonoverlapping test windows did not provide clear evidence of whether shortages increased or decreased. We also examine how execution delays and adjustment limits affect this trade off.

## 1 INTRODUCTION

In cloud computing, a cloud controller uses workload forecasts to request CPU and memory before demand arrives. Too little capacity causes shortages, while too much leaves capacity idle. Under a shared service-level agreement (SLA), a shortage of either resource counts toward the same limit [4, 5, 8, 16, 19]. The controller therefore needs to decide how to divide the allowed shortage risk between CPU and memory.

1 Corresponding author: Yihua Xu, yxu442@gmail.com.

1 Yihua Xu and Jize Li contributed equally and share first authorship.

The way this risk is divided affects how much extra capacity each resource needs. For example, a 5% shared shortage cap can allow 2.5% for CPU and 2.5% for memory, an obvious equal share. If spare memory costs more than spare CPU, the controller may allow more shortage risk for memory and reserve less extra memory. The total risk budget would stay the same, but the capacity requests, shortage frequencies, and surplus costs would be significantly different.

The available forecast-error history impact which splits can be used. We use split conformal prediction [24] to calculate extra capacity from past forecast errors. A smaller shortage allowance needs more past observations to support the calculation, so some splits cannot be used with a short history. However, extensive records may also provide little extra information when errors in nearby periods are similar. A split can work well in one period and poorly in another. Before comparing costs, we check whether the history supports each split and whether it performs consistently across different parts of that history.

With the above information, our main contribution is a joint provisioning rule that checks the risk splits before choosing one by cost. Among the splits that pass, the controller selects the one with the lowest combined cost of shortages and idle capacity on the calibration data. If none passes, it continues provisioning with a calibrated fallback, without assigning the requested SLA label. Figure 1 shows the procedure. Passing the checks allows the controller to use the selected split, but does not ensure that it will meet the SLA target on future workloads.

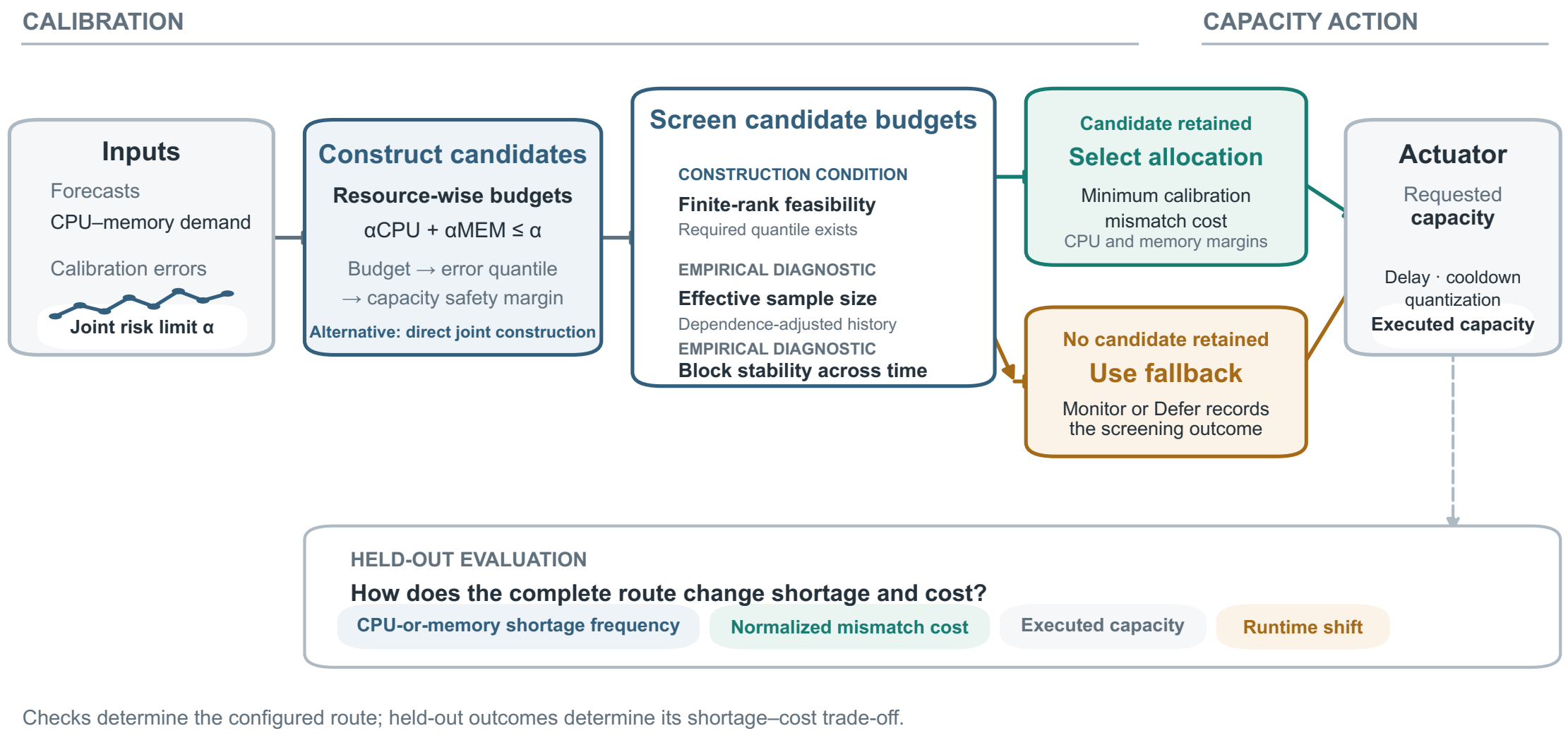

Figure 1: Controller workflow demonstration. Admit executes the selected allocation; Monitor and Defer distinguish why there is no candidate pass and invoke fallbacks. The actuator / executor converts requested capacity into executable capacity. Held-out outcomes evaluate shortages and cost; Adaptive conformal inference (ACI) is a separate comparison method that updates during operation, as described in Section 5.

## 2 RELATED WORK

Cloud resource managers use forecasts and feedback to choose capacity under operational constraints [8, 16, 17, 19, 22]. TIMBRE [13] uses external event reports for time-series forecasting. Newsvendor, predict-then-optimize, and chance-constrained methods frame the shortage-surplus trade-off [4, 5, 9, 15]. Conformal prediction supplies the calibration-based capacity margin used here [18, 24], while studies of dependence and shift motivate attention to temporal structure and online adaptation [6, 12, 26, 27]. Wu et al. [25] use conformalized quantile regression for carbon allowance price intervals.

The closest allocation method is Dynamic Conformal Prediction for Multi-Target Regression (DCP-MT) [20]. It combines one-dimensional conformal predictors into a joint region and selects per-target error rates to reduce region volume under a global error budget. The present controller instead selects CPU and memory budgets by calibration mismatch cost, filters candidates using ESS and block-stability rules, and maps an empty retained set to fallback service. Conformal uncertainty sets for robust optimization [14] connect calibrated regions to downstream decisions rather than directly determining resource requests.

Learn then Test [2] uses valid risk tests and family-wise error control to retain candidate parameters. Conformal Risk Control [3] calibrates a bounded monotone loss, complementing risk-controlling prediction sets [7]. Our cost-based selection uses neither rule and does not inherit their guarantees. Selective prediction [10] can withhold a prediction, and CASEL [28] chooses runtime interventions for tool-using agents. We withhold only the SLA label and provide fallback capacity when no candidate passes.

## 3 PROBLEM FORMULATION

Time is discretized into hourly slots. For resource $j \in \{1, \dots, m\}$, $D_{j,t} \geq 0$ is demand and $C_{j,t} \geq 0$ is provisioned capacity. The shortage and idle surplus are $s_{j,t} = (D_{j,t} - C_{j,t})_+$ and $e_{j,t} = (C_{j,t} - D_{j,t})_+$. With normalized resource scales $g_j > 0$, weights $\omega_j > 0$, and $\sum_j \omega_j = 1$, the mismatch objective is

$$J(C) = \frac{1}{T}\sum_{t=1}^{T}\sum_{j=1}^{m}\omega_j \frac{c_u s_{j,t} + c_o e_{j,t}}{g_j}. \tag{1}$$

A joint shortage occurs when demand exceeds capacity for any protected resource. The controller assigns each resource a risk budget $\alpha_j$, with $\sum_j \alpha_j \leq \alpha_{\text{joint}}$. Each budget selects a calibration-residual rank and therefore a capacity margin. With $n$ aligned observations for $m$ resources, the resource-wise construction requires $n \geq \lceil m/\alpha_{\text{joint}} \rceil - 1$ for finite ranks. This event is a capacity-level SLA proxy, not a latency or queueing guarantee. Equation (2) states the desired target, not a proven property of the selected controller.

$$V_t^{\text{joint}} = 1\{\exists j : D_{j,t} > C_{j,t}\}, \qquad \Pr(V_t^{\text{joint}} = 1) \leq \alpha_{\text{joint}}. \tag{2}$$

The service level is $\beta_{\text{joint}} = 1 - \alpha_{\text{joint}}$. Under ideal execution, provisioned and requested capacity coincide; replay evaluates Eqs. (1) and (2) using executed capacity from Eq. (10).

A causal point forecaster yields $\hat{\mu}_{j,t}$, and calibration residuals are $r_{j,i} = D_{j,i} - \hat{\mu}_{j,i}$. For sorted residuals, one-sided split conformal [24] uses

$$q_j(\alpha_j) = r_{j,(\lceil (1-\alpha_j)(n_j+1) \rceil)}, \qquad C_{j,t} = \max\{0, \hat{\mu}_{j,t} + q_j\}. \tag{3}$$

The rank exists without clipping only when

$$n_j \geq \left\lceil \frac{1}{\alpha_j} \right\rceil - 1. \tag{4}$$

For a resource wise construction, budgets satisfy $\alpha_j > 0$ and $\sum_j \alpha_j \leq \alpha_{\text{joint}}$. Its smallest attainable joint budget is

$$\alpha_{\min}^{\mathrm{rw}} = \sum_{j=1}^{m} \frac{1}{n_j+1}. \tag{5}$$

## 4 EVIDENCE AWARE ADMISSION CONTROLLER

### 4.1 Candidate Checks

Equation (4) gives the finite-rank requirement for split conformal. The controller then applies two empirical checks. Motivated by effective-sample-size diagnostics for dependent observations [11, 23], it substitutes the floor of the count in Eq. (6) into the rank requirement. The autocorrelations are estimated from each resource's calibration residuals. This heuristic substitution is not a finite-sample coverage theorem. Each candidate is also evaluated by leaving out contiguous calibration blocks in turn.

$$n_{\mathrm{eff},j} = \frac{n_j}{1+2\sum_{k=1}^{K_\rho(n_j)} \max(\hat{\rho}_{j,k},0)}, \qquad K_\rho(n) = \min(24, n-2). \tag{6}$$

Calibration is also split into $K_b$ contiguous blocks. A threshold is fitted without block $b$ and evaluated inside it. With $\hat{c}_b$ denoting the resulting coverage, a candidate is stable only if

$$\min_b \hat{c}_b \geq (1-\alpha) - \delta. \tag{7}$$

The primary configuration fixes $K_b = 4$ and $\delta = 0.05$. Resource wise candidates must pass both marginal resource and union event block checks. In Eq. (7), the resource-level check uses that resource's risk budget, while the joint check uses the target for a CPU-or-memory shortage.

The status applies to the candidate set. Defer is returned when no budget has finite-rank support or none of the rank-feasible budgets passes the ESS check. Monitor is returned when at least one passes both checks but none passes all resource and joint block checks. Admit means that at least one candidate passes the configured finite-rank, ESS, and block-stability checks. It permits execution of the selected allocation but does not establish a formal guarantee of its future coverage.

### 4.2 Budget Selection and Fallback

Let $\mathcal{A}_{\mathrm{check}}$ contain the rank feasible budgets retained by the ESS and block checks. The controller selects the budget in $\mathcal{A}_{\mathrm{check}}$ with the lowest calibration mismatch cost and uses its resource quantiles to form requested capacity. If the set is empty, it executes the calibrated Static or Reactive fallback. Algorithm 1 gives the check order and tie breakers.

$$\boldsymbol{\alpha}^* \in \arg\min_{\boldsymbol{\alpha}\in\mathcal{A}_{\mathrm{check}}} \hat{J}_{\mathrm{cal}}(\boldsymbol{\alpha}), \tag{8}$$

Under exchangeability, split conformal prediction gives marginal coverage for a given risk budget. The union bound combines valid resource-level bounds. Here, the same calibration data form the bounds, check candidates, estimate costs, and select the budget. The fixed-budget result does not by itself guarantee coverage for this selected budget. We test the complete procedure on held-out windows.

A few call-outs: Admit uses the selected allocation. Monitor and Defer use fallback without the requested SLA label. Algorithm 1 uses only calibration data to select the budget.

### 4.3 Evidence Acquisition

After Defer, the controller reports the additional aligned observations needed for nominal rank feasibility and an ESS-based projection that holds the observed $n_{\text{eff}}/n$ ratio fixed.

It also reports whether relaxing the target or switching construction changes nominal feasibility.

$$\beta_{\max}^{\text{rw}} = 1 - \alpha_{\min}^{\text{rw}} = 1 - \sum_j \frac{1}{n_j+1}. \tag{9}$$

Algorithm 1: Evidence aware joint SLA admission controller

**Input:** proper history, calibration residuals, joint target $\alpha$, costs $c_u, c_o$, actuator configuration

1. Enumerate construction specific attainable budgets
2. **if** no finite rank candidate exists then
3. **return** Defer, calibrated fallback, evidence plan
4. **end if**
5. Apply ESS, resource-level block, and joint block checks
6. **if** no effective size supported candidate exists then
7. **return** Defer, calibrated fallback, collect/relax/switch options
8. **end if**
9. **if no supported candidate passes all block checks then**
10. **return** Monitor, calibrated fallback, additional monitoring
11. **end if**
12. Select $\boldsymbol{\alpha}^*$ by Eq. (8)
13. Break cost ties by lower calibration joint violation, then lower maximum resource violation, lower normalized capacity, and lexicographic budget order.
14. Form the requested capacity from Eq. (3)
15. **return** Admit, selected risk budget, requested capacity

### 4.4 Actuator Model

The actuator rounds requested capacity $C_t^{\text{req}}$ upward, applies it after delay $d$, and enforces cooldown $h$. With quantizer $Q_\Delta$,

$$C_t^{\text{act}} = Q_\Delta(C_{t-d}^{\text{req}}) \quad \text{when } t - t_{\text{last}} > h, \tag{10}$$

otherwise capacity is held. Replay cost adds $c_s \sum_t \|C_t^{\text{act}} - C_{t-1}^{\text{act}}\|_1$ after normalization.

For each later change $k$, latency is $L_k^{\text{act}} = \tau_k - s_k$. Paired signed cost delta is $\Delta J^{\text{act}} = J^{\text{act}} - J^{\text{ideal}}$ and may be negative on a realized path. In this case, the request time precedes the application time. Then, we measure command to effect latency. The initial capacity corresponds to the first causal request which rounded upward. All later requests are quantized and delayed, and a change is applied only when the cooldown condition in Eq. (10) is met.

## 5 EXPERIMENTAL DESIGN

Table 1 specifies the primary controlled design. Following burn-in, we have a sequence of: proper train, calibrate and outer holdout per seed. We estimate log AR(1) with intercept and single lag, run OLS on proper train data (separately for CPU and memory). The resulting coefficients and scale factors are then kept constant. Predictions are made from all prior timepoints up to and including any outer values seen so far in the causal setting. No information from current or subsequent outer timepoints informs budget allocation, diagnostics etc.

Table 1: Primary controlled configuration. All tuning precedes outer-window evaluation.

| **Component** | **Frozen primary setting** |
|---|---|
| History and replication | 500-hour burn-in; 1,000 proper-training hours; 48, 96, 192, or 384 calibration hours; 240 outer hours; 100 independent seeds per cell. |
| Demand process | Bivariate log AR(1), means log(100) and log(150); innovation variances 0.0144 and 0.0064, covariance 0.0048. S1: coefficient 0.10, Gaussian innovations. S4: the same calibration law; CPU ×1.15 and memory ×1.25 only in the outer window. |
| Targets and candidates | Primary joint service level 95%; sensitivities 90% and 99%. Resource budgets $k/(n+1)$, $k = 1,\dots,n$, are enumerated subject to their sum not exceeding the joint risk limit. |
| Empirical checks | ESS lag limit $\min(24,n-2)$, positive autocorrelation sum, floored effective count; four contiguous leave-one-block-out blocks; tolerance 0.05. |
| Mismatch cost | Shortage coefficient 5; surplus coefficient 1. Resource scales are proper-training mean demands. CPU and memory weights are 0.5 each. Cost is averaged over hours. |
| Constrained replay | Delay 3 hours, cooldown 6 hours, upward quantum 5% of each proper-training mean. The switching coefficient is 0.05. Ideal replay removes these actuation constraints but retains switching in total cost. |
| Uncertainty | 2,000 paired percentile bootstrap resamples; 95% intervals. Independent seeds are the simulation resampling unit, not hours; both policies use the same resampled seeds. |

Static sets each resource's capacity to a proper-training demand quantile. Reactive multiplies the preceding observed demand by one plus a buffer. Calibration evaluates the Cartesian products of 103 quantiles per resource (100 equally spaced from 0.50 to 0.995, plus 0.997, 0.999, 0.9995) and 104 buffers (101 from 0 to 1, plus 1.25, 1.50, 2.0). Fallback minimizes calibration mismatch cost among candidates meeting the calibration joint-shortage proxy; if none qualifies, it minimizes violation before cost. Algorithm 1 gives allocation tie-breaking.

Full Controller selects from the retained budget set and otherwise executes fallback. Fallback Only always executes its calibration-selected Static or Reactive policy. Admission-off retains finite-rank feasibility, the same cost objective and tie-breaking, but omits ESS and block checks. It allocates in every reported S1 context. Allocation-only comparators use equal resource budgets or a fixed 25% CPU / 75% memory split. Direct joint conformal calibrates the maximum, across resources, of the residual divided by its proper-training mean; its scalar quantile supplies a scale-proportional margin to each resource. It has its own rank, ESS, and joint-block checks. Table 3 conditions all four allocations on the same eligible seeds, without fallback.

Table 6 uses a separate synthetic log-AR(1) process for 2, 4, or 8 generic resources, not additional measured resource types. Its temporal coefficient is 0.30; resource log-location parameters and innovation scales are evenly spaced from 4.60 to 5.05 and from 0.08 to 0.12. Standardized Gaussian, Student's t with five degrees of freedom, or centered-lognormal innovations (log-scale 0.60) are mixed using an equicorrelation Cholesky factor. The primary slice uses Gaussian innovations, cross-resource correlation 0.35, and 100 seeds at the first tested rank-feasible calibration length. The exploratory grid uses correlations -0.10, 0, 0.35, and 0.75, all three innovation families, and 30 seeds per cell; Table 6 reports its 95% target and 384-hour slice. Burn-in, proper-training, and outer-window lengths remain 500, 1,000, and 240 hours. Resource weights are equal. Structural fallback tuning uses one shared quantile or buffer across resources, omitting 0.9995 from the Static grid. Candidate sets are enumerated up to 4,096 budgets; larger sets use a reproducible subset containing balanced and sampled allocations, so reported admission is relative to the evaluated set.

The external evaluation uses the 500-VM random-collection subset of Bitbrains GWA-T-12 [21]. For each VM, we average CPU usage in MHz and memory usage in KB within each hour, then sum across VMs. The resulting 2,136 observed hours form three contiguous segments; missing hours are not filled. Within a segment, proper training expands from its start with at least 144 hours, followed by 384 calibration hours and a 96-hour outer window, advancing by 96 hours. The five primary outer windows do not overlap, but adjacent calibration windows overlap by 288 hours and share earlier training history. A secondary panel ranks VM identifiers by SHA-256 of a fixed salt and identifier, then assigns them round-robin to ten disjoint pools of 50 VMs before computing outcomes. The same gap-safe window rule yields 48 paired pool-windows: nine pools contribute five windows each and one contributes three, spanning five shared outer-start times. This panel is not an independent dataset.

Forecaster sensitivity uses fixed 24-hour seasonal-naive and linear log-lag-1-and-24 predictors; no forecaster is selected using outer outcomes. Adaptive conformal inference (ACI) [12] is a separate runtime comparator. It starts from the requested target. After each request, it adds 0.05 times the difference between the nominal target and the observed binary joint-shortage indicator to its current level, then clips to [0.0001, 0.5]. Its joint-score history expands after the decision; the update affects the next request. It clips its runtime quantile rank to the available history and is not a finite-rank guarantee. Admission and budget selection do not consume this feedback.

An admitted-window exceedance means that an admitted context's realized outer-window rate exceeds its target; its count is divided by all contexts and is zero when none is admitted. All-context shortage and cost include fallback execution. Differences are Full Controller minus comparator, with rate differences in percentage points (pp). Confidence intervals (CIs) use paired resampling; Table 3 uses the same 68 seeds for all comparisons. Bitbrains intervals resample windows, with source-segment sensitivity. For the secondary panel, hierarchical bootstrap samples pools and then windows within sampled pools; two-way cluster bootstrap independently samples pool and outer-start clusters and weights each paired difference by the product of their multiplicities. Both use 2,000 resamples and percentile 95% intervals with base random seed 20260821; the 48 paired windows are not treated as independent replications. Capacity changes count later applied capacity-vector changes; switching volume averages resource-scaled, weighted absolute changes over hours. Replay total cost adds the switching term to executed mismatch cost.

## 6 RESULTS

### 6.1 Full Controller versus Fallback

Full Controller and Fallback Only were identical at 48 and 96 calibration hours because every seed used fallback. At 192 hours, across all 100 seeds, the full controller reduced CPU-or-memory shortage frequency by 1.04 percentage points and increased normalized mismatch cost by 0.0091 relative to fallback only. The same trade-off remained at 384 hours with a smaller difference. Table 2 reports admission rates, absolute outcomes, paired intervals, and both calibration lengths.

Table 2: Primary S1 routing comparison at the 95% joint target. Differences are Full Controller minus Fallback Only; 95% paired bootstrap intervals use 100 common seeds. Shortage differences are percentage points (pp); cost is normalized mismatch cost. Means and paired differences are rounded independently.

**Panel A. Admission and all-context shortage**

| Calibration hours | Admit rate | Fallback shortage | Full shortage | Paired Δ shortage [95% CI], pp |
|---|---|---|---|---|
| 192 | 86% | 5.80% | 4.77% | −1.04 [−1.36, −0.75] |
| 384 | 100% | 5.59% | 4.95% | −0.64 [−0.87, −0.42] |

**Panel B. All-context cost**

| Calibration hours | Fallback cost | Full cost | Paired Δ cost [95% CI] |
|---|---|---|---|
| 192 | 0.2134 | 0.2225 | +0.0091 [+0.0073, +0.0109] |
| 384 | 0.2136 | 0.2190 | +0.0054 [+0.0044, +0.0064] |

### 6.2 Allocation on Common Eligible Seeds

On the 68 seeds where all four allocations were eligible, the selected allocation had lower mismatch cost than direct joint, equal resource-wise, and fixed 25/75 allocation. The shortage-frequency difference from direct joint conformal was inconclusive. Shortage frequency was higher than under the equal and fixed allocations. This common-support comparison identifies a cost–shortage trade-off among allocations rather than a complete-controller effect.

Table 3: Common-support allocation comparison in S1 at 95% and 384 calibration hours. Differences are selected allocation minus comparator on the same 68 eligible seeds. Intervals are 95% paired percentile bootstrap intervals over independent seeds.

| Comparator | n | Paired Δ mismatch cost [95% CI] | Paired Δ shortage [95% CI], pp |
|---|---|---|---|
| Direct joint conformal | 68 | −0.0129 [−0.0151, −0.0108] | +0.21 [−0.09, +0.51] |
| Equal resource-wise | 68 | −0.0077 [−0.0088, −0.0065] | +0.45 [+0.28, +0.64] |
| Fixed 25/75 | 68 | −0.0258 [−0.0286, −0.0230] | +0.32 [+0.02, +0.61] |

### 6.3 Admission Off Ablation and Diagnostic Sensitivity

At 48 calibration hours, the controller admitted no contexts and therefore produced no admitted-window target exceedances. It served every context with fallback, which lowered mismatch cost but increased all-context shortage frequency relative to admission off. Admission rates depend on the diagnostic settings. Across the tested lag and block-tolerance settings, admission ranged from 0% to 49% at 96 calibration hours and from 67% to 100% at 192 hours.

Table 4: S1 admission-off comparison at the 95% target on 100 common seeds. Admission-off checks finite rank only. Panel A counts allocations carrying the SLA label and, separately, those also exceeding the target in their outer window; every denominator is all 100 contexts, not the admitted subset. Panel B reports all-context Full Controller minus admission-off differences with 95% paired bootstrap intervals.

**Panel A. Labeled allocations and outer-window exceedances**

| Calibration hours | Admission-off labeled / all | Full admitted / all | Admission-off labeled and exceeding / all | Full admitted and exceeding / all |
|---|---|---|---|---|
| 48 | 100/100 | 0/100 | 28/100 | 0/100 |
| 96 | 100/100 | 0/100 | 38/100 | 0/100 |
| 192 | 100/100 | 86/100 | 39/100 | 29/100 |
| 384 | 100/100 | 100/100 | 40/100 | 41/100 |

**Panel B. All-context service outcomes**

| Calibration hours | Paired Δ shortage [95% CI], pp | Paired Δ mismatch cost [95% CI] |
|---|---|---|
| 48 | +4.55 [+3.89, +5.25] | −0.0446 [−0.0525, −0.0376] |
| 96 | +1.95 [+1.62, +2.29] | −0.0165 [−0.0192, −0.0141] |
| 192 | −0.06 [−0.34, +0.17] | +0.0018 [+0.0007, +0.0028] |
| 384 | −0.01 [−0.08, +0.05] | +0.0007 [+0.0003, +0.0012] |

### 6.4 Bitbrains Results and Forecaster Sensitivity

In the five-window Bitbrains analysis, Full Controller had a lower average shortage frequency than Fallback Only, but the 95% interval ranged from a 6.04-point reduction to a 0.42-point increase. Panel B of Table 5 reports a -1.89 pp shortage difference in the secondary 48-pair panel; both bootstrap shortage intervals exclude zero, while the two-way clustered cost interval includes zero. Shared time periods prevent treating these windows as independent replications or replacing the inconclusive primary five-window result. Forecaster sensitivity mainly changed which windows reached Admit. In the seasonal-naive comparison, the four non-admitted windows execute the same fallback on both arms; the remaining window also has an exactly zero shortage-rate difference. Thus the [0.00, 0.00] interval in Panel A of Table 5 reflects the observed paired differences, not evidence of population equivalence.

Table 5: Bitbrains at 95% and 384 calibration hours. Differences are Full Controller minus Fallback Only. Panel A resamples five nonoverlapping outer windows with shared calibration history; a zero interval means all five observed paired shortage differences are zero, not equivalence. Panel B uses log AR(1) and 48 paired pool-windows from ten pools and five shared outer-start times. Its 95% intervals use hierarchical or two-way cluster bootstrap; cost is normalized mismatch cost.

**Panel A. Primary five-window analysis**

| Forecaster | Admitted windows | Δ shortage [95% CI], pp | Δ mismatch cost [95% CI] |
|---|---|---|---|
| Log AR(1) | 5/5 | -2.50 [-6.04, 0.42] | -0.076 [-0.313, 0.083] |
| Seasonal naive | 1/5 | 0.00 [0.00, 0.00] | 0.049 [0.000, 0.148] |
| Linear lag | 4/5 | -2.50 [-5.62, -0.21] | -0.088 [-0.336, 0.077] |

**Panel B. Secondary 48-pair pool analysis**

| Interval method | Δ shortage [95% CI], pp | Δ mismatch cost [95% CI] |
|---|---|---|
| Hierarchical bootstrap | -1.89 [-3.00, -0.87] | -0.141 [-0.248, -0.046] |
| Two-way cluster bootstrap | -1.89 [-4.08, -0.40] | -0.141 [-0.373, +0.043] |

### 6.5 Resource-Dimension Feasibility Frontier

Dividing the joint risk limit across more resources raises the nominal history requirement. Under the configured synthetic stress grid, two- and four-resource cells admitted throughout the reported 384-hour slice, whereas the eight-resource exploratory cells admitted only about half of contexts. Table 6 reports the exact rank frontiers, seed counts, and the range of admission rates across cells; the latter is not a confidence interval.

Table 6: Resource dimension at the 95% joint target under the synthetic process in Section 5. Panel A uses Gaussian innovations, cross-resource correlation 0.35, and 100 seeds at the first tested rank-feasible length. Panel B averages 12 correlation-by-innovation cells with 30 seeds each at 384 hours; brackets give their minimum and maximum admission rates, not a confidence interval. Eight-resource cells in Panel B evaluate a 4,096-budget subset.

**Panel A. Exact frontier and primary 100-seed slices**

| Resources | Exact rank frontier | Primary calibration hours | Primary seeds | Primary admission |
|---|---|---|---|---|
| 2 | 39 | 48 | 100 | 0/100; Defer (ESS) |
| 4 | 79 | 96 | 100 | 0/100; Defer (ESS) |
| 8 | 159 | 192 | 100 | 0/100; Defer (ESS) |

**Panel B. Exploratory 30-seed cells at 384 hours**

| Resources | Exploratory seeds per cell | 384-hour admission [range] |
|---|---|---|
| 2 | 30 | 100.0% [100%, 100%] |
| 4 | 30 | 100.0% [100%, 100%] |
| 8 | 30 | 56.4% [47%, 63%] |

### 6.6 Actuation Constraints Attenuate the Policy Contrast

With a 3-hour delay, 6-hour cooldown, and 5% quantization, the S1 shortage difference between Full Controller and Fallback Only narrowed from 0.64 to 0.11 percentage points, while Full Controller retained a higher cost. After the S4 shift, both predeployment routes had high absolute shortage frequencies; ACI and Reactive, which update from recent outcomes, had lower reported shortage frequencies under the same actuator. Tables 7 and 8 report paired intervals and absolute outcomes.

Table 7: S1 execution results with a 95% joint SLA target and 384 hours of calibration data. The comparison uses the same 100 simulation seeds for both methods and reports Full Controller minus Fallback Only differences with 95% bootstrap intervals. For the full controller, latency measures how long a requested capacity change takes to take effect, rather than application response time. Total cost combines shortage, idle-capacity, and switching costs.

| Actuator delay / cooldown / quantum | Paired Δ shortage [95% CI], pp | Paired Δ total cost [95% CI] | Mean latency (h) |
|---|---|---|---|
| Ideal | −0.64 [−0.87, −0.42] | +0.0059 [+0.0049, +0.0069] | 0.0 |
| 3 h / 6 h / 5% | −0.11 [−0.35, +0.09] | +0.0050 [+0.0028, +0.0072] | 3.0 |

Table 8: S4 outcomes at 95%, 384 calibration hours, and the same 3 h / 6 h / 5% actuator. Shortage and total cost show means with 95% seed-level bootstrap intervals; changes count mean applied capacity-vector changes. Each row uses 100 seeds.

| Method | Shortage [95% CI], % | Total cost [95% CI] | Changes |
|---|---|---|---|
| Full Controller | 62.74 [61.02, 64.30] | 0.2715 [0.2656, 0.2772] | 25.3 |
| Fallback Only | 72.28 [70.73, 73.78] | 0.3090 [0.3010, 0.3175] | 0.0 |
| ACI | 7.90 [7.59, 8.19] | 0.2786 [0.2748, 0.2826] | 33.0 |
| Reactive | 6.64 [6.17, 7.12] | 0.3815 [0.3766, 0.3864] | 34.0 |

### 6.7 Actions for Short Calibration Histories

As seen above, there are insufficient aligned data points to construct finite ranks for the 95% target under a resource-wise approach in the 24-hour Alibaba case. Defer uses a fallback method and returns three recommendations: gather further historical data, decrease the service-level target, or consider direct joint construction. Table 9 shows the additional history needed under the nominal and ESS requirements. For Alibaba 2018 [1], we take the mean of CPU and RAM usage per machine-hour and aggregate over all machines. We compute this over four rolling windows with expanding proper-training data, 24 aligned calibration hours, and 24 test hours. Any window containing an empty source hour during calibration, the 24-hour causal lookback, or the test period is dropped.

Table 9: Alibaba case with a 95% target and 24 aligned calibration hours. The ESS gap is a planning estimate for this case, assuming the observed ESS-to-history ratio stays constant. The relaxed target meets the rank requirement only. The ESS gap is reported as the median across four rolling windows.

| Option | Reported requirement or outcome |
|---|---|
| Collect for nominal rank | 15 additional aligned hours |
| Collect under ESS projection | 104 additional hours (median) |
| Relax resource-wise target | 92% nominal rank-attainable service level |
| Switch to direct joint construction | Rank-constructible; separate diagnostics required |
| Defer action | Calibrated fallback without the SLA label |

## 7 DISCUSSION

### 7.1 Interpretation and Operating Scope

When at least one candidate is retained, the experiments compare the resulting shortage frequency and mismatch cost with fallback. Whether a measured shortage reduction warrants the additional cost depends on the operator's cost preferences. When no candidate is retained, fallback keeps service operational. The zero admitted-window exceedance count then follows from zero admission, so service performance is assessed from all-context outcomes.

With this combination of delay, cooldown, and quantization, we saw a smaller difference in shortage frequency between Full Controller and Fallback Only. Hence, allocation performance should be measured using executed rather than requested capacity. However, our tests do not show which actuator constraint caused the reduction.

In S4, Full Controller and Fallback Only had high shortage rates after the shift. ACI and Reactive had lower shortage rates because they updated using new data.

### 7.2 Deployment Use

In S1 with 384 calibration hours, the controller evaluated 171 budgets per context, with a median decision time of 14.23 ms. In the larger structural sweep, the p95 decision time was 140.9 ms after pruning. These timings exclude forecasting and cluster latency.

### 7.3 Limitations

The full controller were evaluated on reserved simulated and real-world trajectory data. We cannot guarantee that the final selected risk budget will meet coverage requirements. ESS and block checks were only used to screen candidate risk budgets. We used the same batch of calibration data to screen and select risk budgets. Access to longer term, independent calibration and deployment data would help further validate this approach.

In external validation, we used five main test windows from the Bitbrains dataset. Larger pool panels have shared time-related effects, and the admission depends on the predictive model and diagnostic parameters used. Longer-term, independent multi-resource runtime data would provide stronger evidence.

Our simulated execution model considers delay, cooldown time, capacity quantization adjustments, and capacity switching, but does not account for resource placement, queuing, migration, resource contention, non-linear pricing, or telemetry failure. Resource shortage events serve only as proxy indicators for capacity level SLAs and do not fully represent actual service level agreements. The cost function does not represent actual billing amounts or contract breach compensation. Finally, checks based on calibration data cannot automatically adapt to subsequent changes in data distribution.

## 8 CONCLUSION

In this work, we detail the process of identifying suitable CPU and memory risk budgets through our screening and deploying the cheapest valid option (or Fallback if no options are identified). For both 192h and 384h S1 runs, the complete system showed a reduction in CPU or Memory shortage frequency vs Fallback Only, at the expense of high mismatch cost. We obtained an equivocal result using 5 windows on Bitbrains, although some effect of the Actuation restrictions relative to Fallback was apparent. Overall, these experiments provide guidance as to where this technique could apply and suggest further investigation around actual run capacity and future load.

**Reproducibility and Data Availability.**

The study uses the public Alibaba and Bitbrains traces cited in References [1] and [21]. The scripts and instructions for reproducing the controlled S1 simulation and the paired analysis in Table 3 are available in the https://github.com/jizeli-hub/scp-datacenter-reproducibility.